\documentclass[aps,prl,preprint]{revtex4}
\usepackage{graphicx}
\usepackage{epsfig}
\begin{document}

%%%%%%%%%%%%%%%%%%%%%%%%%%%%%%%%%%%%%%%%%%%%%%%%%%%%%%%%%%%%%%%%%%%%%

\title{Magnetization and Spin Excitations of Non-Abelian Quantum Hall States}

\author{Kun Yang$^{1}$}
\author{E. H. Rezayi$^2$}

\affiliation{$^1$NHMFL
and Department of Physics, Florida State University, Tallahassee,
Florida 32306, USA}
\affiliation{$^2$Physics Department, California State University Los
Angeles, Los Angeles, California 90032, USA}

\date{\today}

\begin{abstract}

Significant insights into non-Abelian quantum Hall states were obtained from studying special
multi-particle interaction Hamiltonians, whose unique ground states are the Moore-Read and Read-Rezayi states for the
case of spinless electrons. We generalize this approach to include the electronic spin-1/2 degree of freedom. We
demonstrate that in the absence of Zeeman splitting the ground states of such Hamiltonians have large degeneracies and
very rich spin structures. The spin structure of the ground states and low-energy excitations can be understood based
on an emergent SU(3) symmetry for the case corresponding to the Moore-Read state. These states with different spin
quantum numbers represent non-Abelian quantum Hall states with different magnetizations, whose quasi-hole properties
are likely to be similar to those of their spin polarized counterparts.

\end{abstract}

\maketitle

%{\it Introduction.}
The possibility of quantum Hall states with
fractionally charged quasiparticles that obey non-Abelian statistics
has attracted tremendous interest recently\cite{day,collins,nayak08},
partly because of the potential of using these non-Abelian
quasiparticles for quantum information storage and processing in an
intrinsically fault-tolerant
fashion~\cite{kitaev03,freedman02,dassarma05,bonesteel05}.
Among such non-Abelian quantum Hall states, the most studied are
the Moore-Read (MR) state\cite{moore91}, which may have been
realized at Landau level (LL) filling factor
$\nu=5/2$\cite{willett87}, and the Read-Rezayi (RR) states\cite{rr},
which may have been realized at $\nu=12/5$\cite{xia} for the case of
level $k=3$ (see below for a definition). In these states the spins
of the electrons occupying the valence Landau level (which in
experimental systems is the first excited Landau level) are assumed
to be fully polarized. However this is an assumption which has not
been fully tested numerically. The only exception is for the
case of $\nu=5/2$ where Morf\cite{morf} showed that the fully
polarized state (which has a large overlap with the MR state) has
lower energy than the spin singlet state, consistent with a more recent work using Monte Carlo to evaluate the energies
of the MR and spin-unpolarized 331 states\cite{dimov}; all other numerical
studies\cite{rh,wan06,rezayi06} {\em assume} full polarization.
%Experimentally, attempts to measure the spin polarization have just started\cite{tracy}.
This is very unsatisfactory because, in typical systems, the Zeeman splitting due to electron spin is smaller than the
Coulomb energy scale by about two orders of magnitude. The situation started to change only very recently since Feiguin
{\em et al.}\cite{feiguin} carefully studied the magnetization of a half-filled first excited LL and found compelling
evidence that suggests the electron spins are fully polarized for the case of Coulomb interaction, even in the {\em
absence} of Zeeman splitting. Experimentally, attempts to detect spin polarization at $\nu=5/2$ are on-going and remain inconclusive at this point\cite{dean}.

In the present paper we take an approach that is different but complementary to that of Ref. \onlinecite{feiguin} and
study the
case of a special 3-body interaction\cite{greiter} that makes the MR
state the unique ground state for spin-polarized electrons at half filling.
The special properties of this interaction allow us to establish a number of exact results. When
applied to the case of spin-1/2 electrons (without Zeeman
splitting), we find that a large ground state degeneracy appears
with {\em different} total spin quantum numbers. These degenerate
ground states are constructed explicitly and they form a single
SU(3) multiplet. Such constructions can be generalized to the RR
states when spin is included. This suggests that this family of non-Abelian quantum Hall
states may have very rich spin structure. We further present numerical evidence suggesting that the low-energy spectrum
of the system is consistent with an emergent SU(3) symmetry in the long-wavelength and low-energy limit for the MR
case.%Our results do indeed point to the possibility of non-Abelian quantum Hall states with rich spin structures.

The 3-body interaction that makes the MR state the exact ground state at half-filling takes the form:
\begin{equation}
 H_{3B} = \sum_{i < j <
k}S_{ijk}[\nabla^2_i\nabla^4_j \delta({\bf r}_i - {\bf r}_j)
\delta({\bf r}_i -{\bf r}_k)], \label{H3B}
\end{equation}
where $S$ is a symmetrizer:
$S_{123}[f_{123}]=f_{123}+f_{231}+f_{312}$, and $f$ is symmetric in its first
two indices. For spinless (or
%equivalently,
spin-polarized) electrons, the following MR state is
the unique zero-energy ground state at half-filling:
\begin{equation}
\psi_{MR}=\left[\prod_{i<j}(z_i-z_j)^2\right]A\left({1\over
z_1-z_2}\cdots{1\over z_{2N-1}-z_{2N}}\right), \label{MR}
\end{equation}
where $A$ is the antisymmetrizer, $N$ is the number of pairs (so we
have $N_e=2N$ electrons), and we neglected the common exponential factor
of LL wave functions. $\psi_{MR}$ is annihilated by $H_{3B}$ because
it vanishes sufficiently fast as three particle coordinates approach
each other. We now generalize $\psi_{MR}$ to include spin degrees of
freedom and construct the following zero energy states in which we
keep the Jastrow factor $[\prod_{i<j}(z_i-z_j)^2]$ of Eq. (\ref{MR})
while we modify the Pfaffian factor $A(\cdots)$:
\begin{equation}
\psi(z_1,\chi_1; \cdots; z_{2N}, \chi_{2N})=
\left[\prod_{i<j}(z_i-z_j)^2\right] A\left[\left({1\over z_1-z_2}\cdots{1\over
z_{2N-1}-z_{2N}}\right)\left(\sum_{\{\chi\}}c_{\{\chi\}}\chi_{12}\cdots\chi_{2N-1,2N}\right)
\right], \label{spin}
\end{equation}
where $\chi_{i}$ is the spin wave function of electron $i$ and
$\chi_{ij}$ is the spin wave function of the pair made up of electrons
$i$ and $j$. Obviously (\ref{spin}) reduces to (\ref{MR}) when we
take $\chi_{ij}=|\uparrow\rangle_i|\uparrow\rangle_j$, so that the
electron spins are fully polarized. Also because the orbital part of
(\ref{spin}) has the same asymptotic behavior as (\ref{MR}) when 3
electrons approach each other, (\ref{spin}) is also annihilated by
$H_{3B}$.

We now consider the constraint on $c_{\{\chi\}}$ imposed by the
antisymmetrizer $A$. Because of the fact that the orbital part is
antisymmetric under the exchange between $z_{2j-1}$ and $z_{2j}$,
$\chi_{2j-1,2j}$ must be {\em symmetric} under such exchange; {\em
i.e.}, $\chi_{2j-1,2j}$ must represent a triplet state formed by
electrons $2j-1$ and $2j$. Furthermore, $c_{\{\chi\}}$ must be
symmetric under the exchange of different pairs $(2j-1, 2j)$ and
$(2k-1,2k)$; as a result $c_{\{\chi\}}$ represents a {\em totally
symmetric} spin state formed by $N$ spin-1 objects. For $N$ spin-1/2
objects, the totally symmetric combination forms a unique
$S_{tot}=N/2$ (or fully-polarized ferromagnetic state) with a
degeneracy of $2S_{tot}+1=N+1$ associated with different $S^z_{tot}$
quantum numbers. For a spin-1 object, on the other hand, $S_{tot}$ is
no longer unique for the totally symmetric combination; it was found that\cite{lieb}
\begin{equation}
S_{tot}=N, N-2, N-4, \cdots, \label{lieb}
\end{equation}
with each value appearing exactly once. The total degeneracy is
\begin{equation}
D_0=\sum_{S_{tot}}(2S_{tot}+1)=(N+1)(N+2)/2.
\label{degeneracy}
\end{equation}
An easier way to understand this larger degeneracy is to recognize
that for each spin-1 object there are 3 internal states associated
with $S^z=0, \pm 1$; thus states formed by totally symmetric
combinations of $N$ spin-1 states form a {\em single} totally
symmetric representation of SU(3)\cite{lieb,reijnders}, which is represented by a row of
$N$ boxes in the Young tableaux or simply the representation $[N]$\cite{georgi}. The result (\ref{lieb})
may be viewed as decomposing a single irreducible representation of
SU(3) into multiple irreducible representations of its subgroup
SU(2).

The result (\ref{lieb}) can also be obtained from an alternative
method.
%which is more readily generalized to the cases corresponding to the RR states at level $k$.
The MR state can also be written as
\begin{equation}
\psi_{MR}=\left[\prod_{i<j\le 2N}(z_i-z_j)\right]S\left[\prod_{0<i<j\le N}(z_i-z_j)^2\prod_{N<k<l\le
2N}(z_k-z_l)^2\right], \label{cluster}
\end{equation}
where $S$ is the symmetrizer. In Eq. (\ref{cluster}) one divides the
electrons into two groups, A and B; within each group one has the
Jastrow factor $\prod_{0<i<j\le N}(z_i-z_j)^2$, which is then symmetrized
among all particles. We now generalize Eq. (\ref{cluster}) to
include electron spins:
\begin{equation}
\psi(z_1,\chi_1; \cdots; z_{2N}, \chi_{2N})=\left[\prod_{i<j\le 2N}(z_i-z_j)\right]S\left[\chi_A\chi_B\prod_{i<j\le N}(z_i-z_j)^2\prod_{N<k<l\le
2N}(z_k-z_l)^2\right], \label{cluster1}
\end{equation}
where $\chi_A$ and $\chi_{B}$ represent the spin wave functions for
clusters A and B respectively. The symmetrization imposes the
following constraints on the spin wave functions: (i) $\chi_A$ and
$\chi_B$ are totally symmetric spin wave functions of $N$ spin-1/2
particles and thus each represents a spin-$N/2$ object; (ii) since
the two clusters are also symmetrized, the total spin is a symmetric
combination of two spin-$N/2$ objects, which leads to Eq.
(\ref{lieb}).

The construction above can be easily extended to the RR
states\cite{rr} at level $k$ to include spin, which are
zero energy states of a special $k+1$-body interaction:
\begin{equation}
\psi(z_1,\chi_1; \cdots; z_{kN}, \chi_{kN})=\left[\prod_{i<j\le kN}(z_i-z_j)\right]
S\left[\prod_{I=1}^k\left(\chi_I\prod_{0<i<j\le
N}(z^{I}_i-z^{I}_j)^2\right)\right], \label{cluster2}
\end{equation}
where we have divided $N_e=kN$ electrons into $k$ clusters, $\chi_I$ is
the spin wave function of the $I$th cluster, and $z^{I}_i$ is the
spatial coordinate of the $i$th electron of the $I$th cluster. Using
the same arguments as before, we find that we have $k$ spin-$N/2$
objects (one from each cluster) forming totally symmetric
combinations; the total ground state degeneracy is
\begin{equation}
D_0={(k+N)!\over k!N!}, \label{degeneracy1}
\end{equation}
which applies to the Laughlin ($k=1$) and MR ($k=2$) cases as well.
It coincides with the totally symmetric $[N]$ representation of the
SU($k+1$) group\cite{georgi}.

Our prediction of the spin quantum numbers for the case of $k=2$ has been
confirmed by exact diagonalization of the 3-body Hamiltonian properly generalized to include spin degrees of freedom, on the sphere for up to
10 electrons. The Hamiltonian of Eq. 1 is not strictly positive definite when
spin reversed states are included.  In addition, it contains an arbitrary scale.
We will work instead with a Hamiltonian made of projection operators:
%\newpage
\begin{eqnarray}
H_{3B}&=&\sum_{m=1}^3 V_m P(3N_{\phi}/2-m,1/2)+\sum_{n=1}^2 V_4^nP_n(3N_{\phi}/2-4,1/2) \nonumber\\
&+&V^\prime_3P(3N_{\phi}/2-3,3/2)+V_5P(3N_{\phi}/2-5,1/2),\label{H3Bproj}
\end{eqnarray}
where $N_\phi$ is the total magnetic flux through the system and $P(L,S)$
projects out the state of angular momentum $L$ and spin $S$. When
such states are not unique we distinguish them with an index $n$. The $V$'s are
the 3-body pseudo-potential parameters\cite{simon} all of which were set to be 1. The projection
operators $P$ have unit eigenvalues as expected. The first 6 terms project out
the states of 3 fermions with relative angular momentum less than 5,
which are absent both in the  MR state and the states of Eq.~
(\ref{spin}) . The last term projects out all 3 fermionic states with relative
angular momentum $m=5$, and spin $S=1/2$\cite{foot1} in which the opposite spins have
relative angular momentum zero, which are also absent in Eq.~(\ref{spin}).

\begin{centering}
\begin{figure}
\epsfig {file=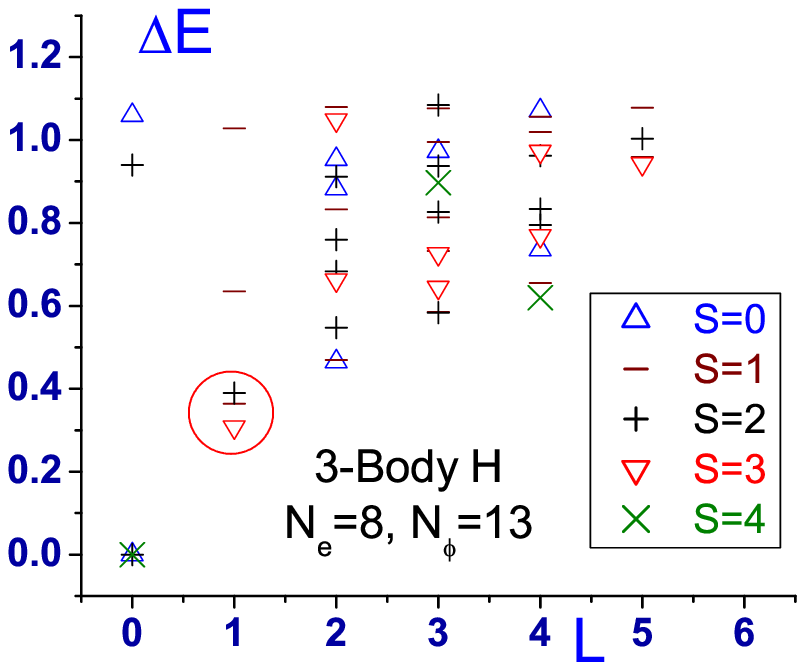,width=80mm}
\epsfig {file=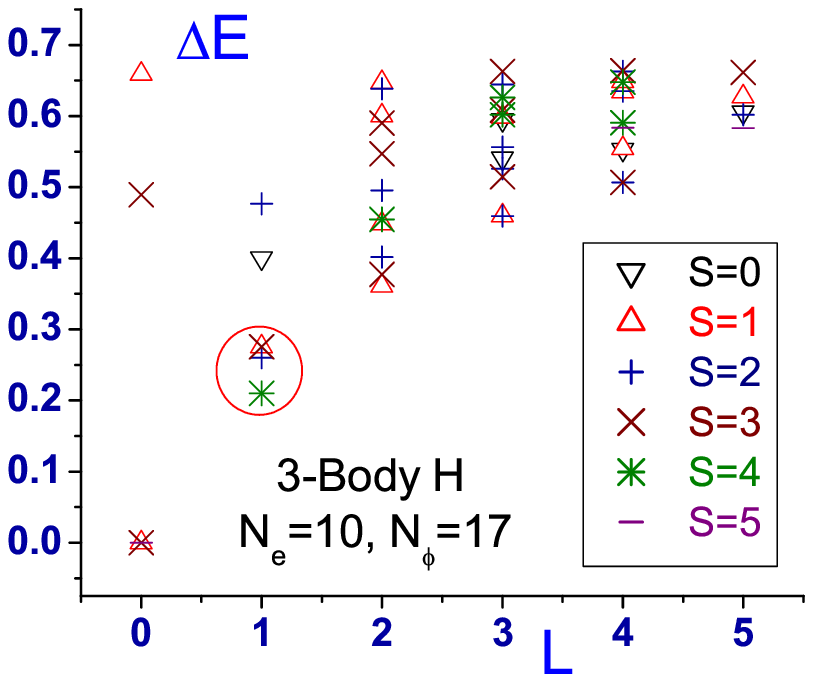,width=80mm}
\caption{(Color online)
Low-energy spectrum of the 3-body Hamiltonian on the sphere at half-filling. The number of flux quanta $N_\Phi$
corresponds to a shift of 3, which is the same as that of the Moore-Read state. The ground states (at $L=0$) with
different total spin quantum number ($S$) form a single totally symmetric SU(3) multiplet corresponding to a
fully-magnetized SU(3) ferromagnet; the low-energy excitations at $L=1$ (inside red circle) are understood to be SU(3)
spinwaves. Upper panel: System with 8 electrons (or 4 pairs); lower panel: 10 electrons or 5 pairs.
}
\label{fig1}
\end{figure}
\end{centering}

Fig. 1 shows the spectra for the cases of 8 electrons (4 pairs) and 10 electrons
(5 pairs) respectively.
We find that the ground states with zero energy all have total
angular momentum $L=0$, which is the same as the MR state, and their spin
quantum numbers indeed take values $S_{tot}=N, N-2, N-4,\cdots$, as predicted.
In addition to ground states, the spectra of the lowest-energy excitations
are also noteworthy. We see in both cases the lowest-energy excited states
have total angular momentum $L=1$ and have spin quantum numbers
$S_{tot}=1,2,\cdots,N-1$, with each multiplet appearing exactly once with
nearly degenerate energies. If the degeneracy were exact, that would result
in a total degeneracy for the first excited states
\begin{equation}
D_1=\sum_{S_{tot}=1}^{N-1}(2S_{tot}+1)=N^2-1.
\label{degeneracy2}
\end{equation}
In the following we argue that this can be understood as the consequence of an emergent SU(3) symmetry at
low-energies.

As discussed above, the ground states can be viewed as a single, totally symmetric SU(3) multiplet. If the system had
an exact SU(3) symmetry, we could view the ground state as a fully-magnetized SU(3) ferromagnet and the SU(3) symmetry
would be spontaneously broken. Then the lowest energy excitations of the system are expected to be SU(3) spin waves.
The lowest-energy spin-wave state would have the smallest possible angular momentum $L=1$ (corresponding to the
smallest momentum in a translationally-invariant system) and has one SU(3) spin ``flipped". In group theoretical
language, a single ``spin flip" means going from the totally symmetric representation $[N]$ to the representation
$[N-1,1]$ (which is represented by two rows in the Young tableaux with $N-1$ and 1 boxes respectively), indicating one
of the SU(3) spins is antisymmetrized with another. This mixed representation indeed has dimension
$N^2-1$\cite{georgi}, in agreement with Eq. (\ref{degeneracy2}), and it is easy to show that when decomposing this
single SU(3) representation into SU(2) representations, one obtains $S_{tot}=1,2,\cdots,N-1$. We thus conjecture the SU(3) symmetry is a property of the Hamiltonian (\ref{H3Bproj}) at low-energy; this is exact for the ground states, but for the excited states it is approximate, and supported by numerical evidence only.
Should the symmetry become asymptotically exact in the long-distance limit, we would expect the degeneracy of
the lowest-energy excited states to improve as system size increases and become asymptotically exact.

In a recent work\cite{dimov}, Dimov {\em et al.} argued that the low-energy effective theory of the ferromagnetic state at $\nu=5/2$ is described by a perturbed CP$^2$ non-linear $\sigma$ model (NL$\sigma$M). The original CP$^2$
NL$\sigma$M possesses SU(3) symmetry; Dimov {\em et al.} argued that, for Coulomb or other generic two-body
interactions, there exist symmetry-breaking perturbations in the effective theory that reduce the SU(3) symmetry down
to SU(2), which is the symmetry possessed by the microscopic Hamiltonian. The 3-body Hamiltonian (\ref{H3Bproj}) also
possesses SU(2) symmetry {\em only}. However our results suggest that for this very special case, the low-energy
physics is very close to the original CP$^2$ NL$\sigma$M with all the symmetry-breaking perturbations vanishing; in
fact it may be possible to tune certain parameters in the Hamiltonian (\ref{H3Bproj}) to reach such a high symmetry
point. If so, such a special 3-body Hamiltonian would be a very useful point of departure for studying the various
possible spin states and low-energy excitations above them at $\nu=5/2$. If the ferromagnetic state at $\nu=5/2$ indeed possesses approximate SU(3) symmetry, it will support {\em two} instead of just one low-energy spin-wave modes, and the skyrmions that appear when $\nu$ deviates from 5/2 will have a richer spin structure\cite{dimov}. Such differences from ordinary SU(2) quantum Hall ferromagnets can be probed using NMR and other experimental methods.

As emphasized earlier, the large spin degeneracies associated with the states described by Eqs. (\ref{spin}) and
(\ref{cluster2}) are special properties of the special multiple-electron interaction Hamiltonians. For a generic
Hamiltonian with SU(2) symmetry, the degeneracy between states in Eqs. (\ref{spin}) and (\ref{cluster2}) with different
$S_{tot}$ will be lifted. They will then represent quantum Hall states with different magnetization that varies
essentially
{\em continuously} from zero to full polarization. In general, one would expect
these states to dominate the magnetization of the system at finite but low
temperatures.
Quasihole excitations on top of these ground states can be constructed in a manner similar to their spin-polarized
counterparts; for example a two-quasihole state on top of the ground state (\ref{spin}) with the same spin quantum
number takes the form
\begin{equation}
\psi_{2qh}=
\left[\prod_{i<j}(z_i-z_j)^2\right]
A\left[\left({(z_1-\eta_1)(z_2-\eta_2)+(z_1-\eta_2)(z_2-\eta_1)\over z_1-z_2}\cdots\right)\left(\sum_{\{\chi\}}c_{\{\chi\}}\chi_{12}\cdots\chi_{2N-1,2N}\right)
\right], \label{2qh}
\end{equation}
where $\eta_1$ and $\eta_2$ are the quasihole coordinates. Multi-quasihole states can be constructed similarly. Just
like the quasihole states of the MR and RR states\cite{readqh}, the locations of the quasiholes do {\em not} uniquely
determine the state when more than two quasiholes are present and the degeneracy grows exponentially with the quasihole
number; these are thus {\em non-Abelian} quasiholes. Their braiding properties may also turn out to be the same as
those of the MR state and will be left to future work.
Another, but less likely\cite{feiguin}, possibility would be a spontaneous
breaking
of the spin $SU(2)$ symmetry that obtains the $SU(3)$ degeneracy for generic
Hamiltonians. If so, the quantum Hall state will be reduced to the 331
Abelian phase.
The 331-state is {\em not} an eigenstate of $S_{tot}$; it can be constructed
as a linear superposition of states of
the form (\ref{spin})\cite{foot2} with different $S_{tot}$ but fixed $S^z_{tot}=0$.

We have benefited from collaboration with Adrian Feiguin, Chetan Nayak and Sankar Das Sarma on a closely related
project and useful conversations with
Steven Simon and Ivailo Dimov. This work was supported by NSF grants No. DMR-0704133 (K.Y.) and DMR-0606566 (E.H.R.).


\begin{thebibliography}{99}

\bibitem{day} C. Day, Phys. Today {\bf 58} (10), 21 (2005).

\bibitem{collins} G. P. Collins, Scientific American {\bf 294} (4), 57 (2006).

\bibitem{nayak08}  Chetan Nayak, Steven H. Simon, Ady Stern, Michael Freedman, and Sankar Das Sarma, arXiv:0707.1889.

\bibitem{kitaev03}
A. Kitaev, Ann. Phys. {\bf 303}, 2 (2003).

\bibitem{freedman02}
M.~H. Freedman, A. Kitaev, and Z. Wang, Commun. Math. Phys. {\bf
227}, 587 (2002); M.~H. Freedman, M. Larsen, and Z. Wang, {\it
ibid.} {\bf 227}, 605 (2002).

\bibitem{dassarma05}
S. Das Sarma, M. Freedman, and C. Nayak, Phys. Rev. Lett. {\bf 94},
166802 (2005).

\bibitem{bonesteel05}
N.~E. Bonesteel, L. Hormozi, and G. Zikos, and S.~H. Simon, Phys.
Rev. Lett. {\bf 95}, 140503 (2005).

\bibitem{moore91}
G. Moore and N. Read, Nucl. Phys. B {\bf 360}, 362 (1991).

\bibitem{willett87}
R.~L. Willett, J.~P. Eisenstein, H.~L. Stormer, D.~C. Tsui, A.~C.
Gossard, and J.~H. English, Phys. Rev. Lett. {\bf 59}, 1779 (1987).

\bibitem{rr} N. Read and E. H. Rezayi, Phys. Rev. B {\bf 59}, 8084
(1999).

\bibitem{xia}
J. S. Xia {\em et al.}, Phys. Rev. Lett. {\bf 93}, 176809 (2004).

\bibitem{morf} R. H. Morf, Phys.
Rev. Lett. {\bf 80}, 1505 (1998).

\bibitem{dimov} Ivailo Dimov, Bertrand I. Halperin, and Chetan Nayak, Phys. Rev. Lett. {\bf 100}, 126804 (2008).

\bibitem{rh} E. H. Rezayi and
F. D. M. Haldane, Phys. Rev. Lett. {\bf 84}, 4685 (2000).

\bibitem{wan06}
X. Wan, K. Yang, and E.~H. Rezayi, Phys. Rev. Lett. {\bf 97}, 256804
(2006);  Xin Wan, Zi-Xiang Hu, E. H. Rezayi, and Kun Yang, Phys. Rev. B {\bf 77}, 165316 (2008).

\bibitem{rezayi06} E.H. Rezayi and N. Read, cond-mat/0608346.

%\bibitem{tracy}  L.A. Tracy, J.P. Eisenstein, L.N. Pfeiffer, K.W. West, March Meeting 07.

\bibitem{feiguin} A.~E.~Feiguin, E. Rezayi, Kun Yang, C. Nayak, and S. Das Sarma, arXiv:0804.4502.

\bibitem{dean} See  C.R. Dean, B.A. Piot, P. Hayden, S. Das Sarma, G. Gervais, L.N. Pfeiffer, and K.W. West,  arXiv:0805.3349
 for the most recent work, and references therein.

\bibitem{greiter} M. Greiter, X.-G. Wen and F. Wilczek, Nucl. Phys.
B {\bf 374}, 567 (1992).

\bibitem{lieb}
E. Eisenberg and E. Lieb, Phys. Rev. Lett. {\bf 89}, 220403 (2002).

\bibitem{reijnders} J.W. Reijnders, F.J.M. van Lankvelt, K. Schoutens, and N. Read, Phys. Rev. Lett. {\bf 89}, 120401
    (2002).

\bibitem{georgi} See, {\em e.g.}, H. Georgi, {\em Lie Algebras in Particle Physics}, 2nd Edition, Perseus Books,
    Reading (1999).

\bibitem{simon}
Steven H. Simon, E. H. Rezayi and Nigel Cooper, Phys. Rev. B {\bf 75}, 195306 (2007).

\bibitem{foot1}
There are only two states of 3 fermions with total angular momentum
$L=3N_\phi/2-5$ and total spin $S=1/2$.

\bibitem{readqh}
N. Read, Phys. Rev. B {\bf 73}, 245334 (2006).

\bibitem{foot2}
The Hamiltomian of Eq. (\ref{H3Bproj}) annihilates the 331-wavefunction and
thus its zero energy eigenstates provide a complete basis for expanding 331.

\end{thebibliography}
\end{document}